# Rapid Reduction of Nitrophenols Using Reusable Magnetic *h*-BN/Ni-NiO Nanocomposites

Anjali Varshney[1], Ritesh Dubey[1], Sushil Kumar[1,*], Tapas Goswami[1,*], and Samar Layek[2,*]

[1]Department of Chemistry, Applied Science Cluster, School of Advanced Engineering, UPES, Dehradun, Uttarakhand 248007, India
[2]Department of Physics, Applied Science Cluster, School of Advanced Engineering, UPES, Dehradun, Uttarakhand 248007, India
* Corresponding authors: sushilvashisth@gmail.com, tapas.t@gmail.com, samarlayek@gmail.com

**Abstract**

The efficient and cost-effective conversion of nitro compounds to amines is crucial for industrial processes and environmental remediation, highlighting the demand for earth abundant metal-based catalysts. In this study, magnetic Ni-NiO nanostructures and their composites with two-dimensional hexagonal boron nitride (*h*-BN) were synthesized *via* a simple, scalable combustion method. The structural, morphological, and compositional properties of the synthesized materials were thoroughly investigated using powder X-ray diffraction (PXRD), scanning electron microscopy (SEM), transmission electron microscopy (TEM), X-ray photoelectron spectroscopy (XPS), and UV-Vis analysis. The catalytic activity of both Ni-NiO and *h*-BN/Ni-NiO nanostructures was evaluated for nitrophenol reduction as model reaction. The *h*-BN/Ni-NiO nanocomposite exhibited significantly enhanced catalytic performance compared to pure Ni-NiO, highlighting the synergistic effects between *h*-BN and the Ni-NiO nanoparticles. Notably, the magnetic nature of the Ni-NiO core facilitated easy recovery of the catalyst using an external magnetic field, and the composite demonstrated excellent stability and reusability upto 6 catalytic cycles with minimal loss of activity. The combination of high catalytic efficiency, magnetic separability, and structural stability enable the *h*-BN/Ni-NiO nanocomposite a promising candidate for green and sustainable catalytic applications, particularly in environmental remediation.

# 1. Introduction

The conversion of nitro compounds to amines is of great significance, as amines serve as fundamental building blocks for a wide range of pharmaceuticals, polymers, dyes, and agricultural products [1–3]. Typically, this chemical transformation requires prolonged reaction times and harsh reaction conditions, leading to low efficiency, limited regioselectivity, and most importantly a substantial environmental impact [4,5]. Nitro compounds, especially nitroaromatics (NACs), are considered hazardous due to their high toxicity. As a result, the United States Environmental Protection Agency (USEPA) has designated them as "Priority Pollutants" requiring their concentrations to be limited to below 10 mg/L prior to discharge into the receiving environment [6,7]. Overexposure to NACs can cause serious health issues, including gastrointestinal infections, neurological disorders, organ damage, and even cancer [8,9]. Therefore, the conversion and/or degradation of these NACs to less toxic species is on high demand. Among the various NACs, *p*-nitrophenol (*p*-NP) is one of the most targeted model substrates which can be easily converted into *p*-aminophenol (*p*-AP), a compound with significantly lower toxicity and valuable industrial applications [10]. In this context, the development of reusable catalysts for rapid, efficient, large-scale, and environmentally benign amination reactions has received increasing attention for both environmental remediation and sustainable industrial processes [11–15].

To facilitate the chemical conversion of nitroaromatics, various catalysts (both homogenous and heterogeneous) have been explored, including nanomaterials, metal-organic frameworks (MOFs), graphene quantum dots, metal/metal oxide nanomaterials, and 2D materials [15–21]. Among these, homogeneous catalysts have been extensively employed for the conversion of nitroaromatics, likely due to their easily accessible catalytic sites in solution [22]. However, they pose challenges in product separation and recycling, particularly when involving noble or toxic metal complexes [23]. In contrast, heterogeneous catalysts, commonly used in industrial processes, offer easier recovery and reusability. But sometimes, they suffer from lower activity and selectivity due to steric hindrance and diffusion limitations that restrict reactant access to catalytic sites [3,23]. To address these limitations, researchers have turned to develop nanoscale metal-based catalysts and supports, which offer enhanced activity and selectivity due to their high surface area and uniform dispersion in solution [24]. The size reduction of the support materials to the nanometer scale allows better emulsion formation and improved porosity, thereby enhances the catalytic efficiency. However, recovering and recycling these catalysts still remains a significant challenge.

Recently, attention has shifted towards using transition metal-based magnetic nanostructures to replace traditional metal/metal oxide nanomaterials. Magnetic nanomaterials have proven to be highly efficient and robust tools in this field due to their ease of synthesis, low cost, and reusability [25]. Additionally, these materials can be easily separated using an external magnetic field, eliminating the need for techniques like liquid-liquid extraction, filtration, and centrifugation. Among these, Ni-based catalysts have gained significant interest in catalytic conversions nitroaromatics due to their excellent magnetic properties, earth abundance, and tunable electronic structure [26–29]. It has been demonstrated that when magnetic nanomaterials are integrated with 2D materials, their properties are significantly enhanced [30–33]. These improved properties of the composites are attributed to high electron mobility, low energy consumption, and a large surface area.

In this study, Ni-NiO-based magnetic nanostructures and their composites with two-dimensional hexagonal boron nitride (*h*-BN) were successfully synthesized using combustion method. The resulting nanomaterials have been explored for their catalytic potential in the reduction of different nitrophenol compounds. Notably, the incorporation of *h*-BN into the Ni-NiO nanostructure significantly enhanced the catalytic performance in the hydrogenation of nitrophenols. The highlights of this study are 3-fold: (i) easy, one-pot and scalable synthesis of first-row transition metal-based h-BN/Ni-NiO composites, accompanied by comprehensive structural characterization; (ii) evaluating the synergistic effect of h-BN and Ni-NiO in the aqueous-phase reduction of nitroaromatic compounds; and (iii) composites' efficiency in rapid catalytic conversion (*ca.* 3 min), magnetic separation, recyclability, and facile recovery of reaction products.

## 2. Experimental Section

### 2.1 Materials and methods

Hexagonal boron nitride (*h*-BN), glycine, sodium borohydride (NaBH$_4$), and Ni(NO$_3$)$_2$·6H$_2$O, were procured from Aldrich while nitroaromatics used for conversion studies were purchased from SRL chemicals. All the materials were used as received without further purification. Powder X-ray diffraction (PXRD) spectra of Ni-NiO nanostructures and *h-BN*/Ni-NiO nanocomposites were recorded on a Bruker X-ray diffractometer (D8 Advance ECO) using Cu-Kα radiation (wavelength = 1.5406 Å) to study the phase formation and crystal structures. Transmission electron microscopy (TEM) was performed using JEOL electron microscope at an operating voltage of 200 kV whereas field emission electron microscopy (FESEM) was recorded using a JEOL (Model-JSM-7800F Prime) electron microscope for morphological characterization of the nanostructures. X-ray photoelectron spectra (XPS) were obtained using a PHI 5000 VersaProbe III instrument using monochromatic aluminum Kα radiation. The nature of functional groups present in as-synthesized nanostructures were explored using a PerkinElmer FTIR spectrometer. Surface properties of these nanostructures were examined using an Anton Paar (Model No.: Autosorb 6100) Brunauer-Emmett-Teller (BET) analyzer. UV-Vis spectra were recorded using a UV-1800 Shimadzu spectrophotometer (Japan).

## 2.2 Synthesis of Ni-NiO and *h*-BN/Ni-NiO nanostructures

Nanostructures of Ni-NiO were prepared following a previously reported solution-based glycine-nitrate combustion method [34]. Briefly, 20 mL of 0.6 M glycine was mixed with 30 mL of 0.3 M aqueous solution of nickel nitrate under continuous stirring at 80 ºC. After 30 minutes, the temperature was raised to 180 ºC. The heating continued until gel formation which leads to combustion of the solution resulting in black, foamy product. The solid residue was collected, ground into fine powder, and annealed at 300 ºC for 4 hours. To synthesize *h*-BN/Ni-NiO nanocomposites, a similar combustion approach was employed with slight modification. In this case, 100.0 mg of *h*-BN was introduced into the nickel nitrate solution before combustion.

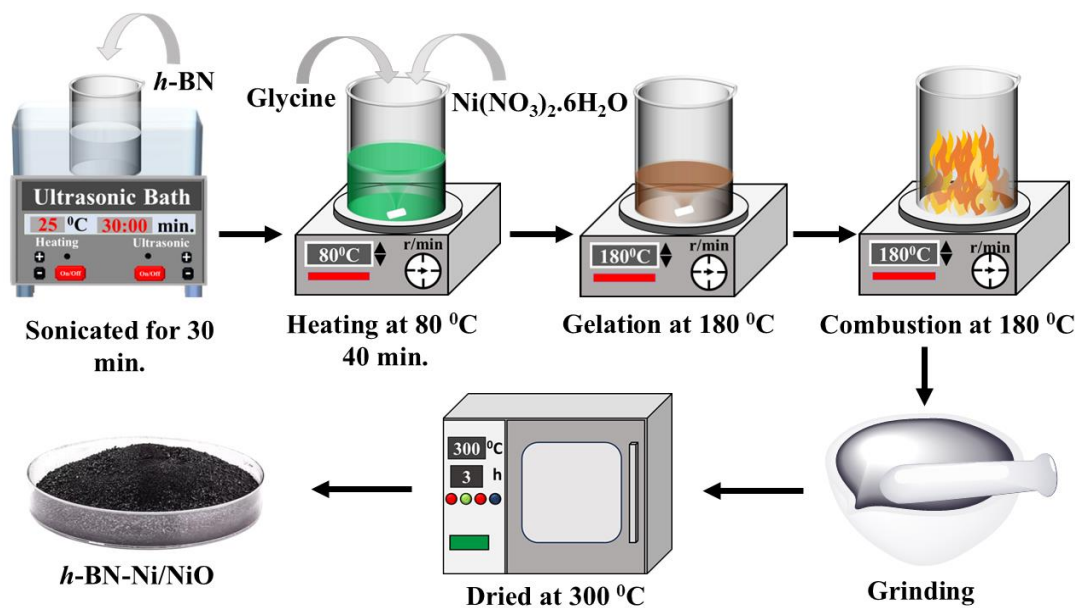

**Scheme 1.** A schematic showing the preparation method used for *h*-BN/Ni-NiO composites.

## 2.3 Evaluation of catalytic activity

The catalytic performance of the as-prepared *h*-BN/Ni-NiO and Ni-NiO nanostructures was evaluated for the reduction of *p*-nitrophenol and *o*-nitrophenol. A 20 mL solution of nitrophenol ($10^{-4}$ M) was prepared and mixed with 7.0 mg of NaBH$_4$. To initiate the reduction reaction, 10.0 mg of the catalyst was added to the above mixture. The reaction was carried out under continuous sonication using a water bath sonicator. The extent of nitrophenol reduction was monitored by recording UV-Vis spectra of the reaction mixture at regular time intervals. The reduction percentage was calculated using the formula: $\% \ Reduction = \frac{A_0 - A_t}{A_0} \times 100$, where $A_0$ is the initial absorbance while $A_t$ stands for absorbance at time t.

## 3. Results and discussion

A straightforward one-pot, solution-based glycine–nitrate combustion method was utilized to synthesize *h*-BN/Ni-NiO nanocomposites (Scheme 1). In this process, nickel nitrate functioned as the oxidizing agent, while glycine served as the fuel. Notably, glycine also played dual roles as a reducing and stabilizing agent during nanostructure formation. Initially, glycine coordinated with Ni(II) ions via its amine (-NH$_2$) and carboxyl (-COOH) groups, forming a Ni(II)-glycine complex [35–37]. Upon heating the homogeneous aqueous solution to 180 °C, gradual water evaporation led to gelation. This gel spontaneously ignited due to a highly exothermic redox reaction between residual nitrate and excess glycine. The resulting combustion generated intense heat and copious gaseous byproducts, producing a porous composite of *h*-BN and nickel nanoparticles. A subsequent calcination step at 300 °C facilitated the surface oxidation of nickel, yielding the final *h*-BN/Ni-NiO nanostructure.

Room temperature PXRD patterns of Ni-NiO nanostructures and *h*-BN/Ni-NiO nanocomposites are shown in Fig. 1(a)-(b). Intense diffraction peaks at 2$\theta$ values 37.19°, 43.21°, 62.79°, 75.31° and 79.30° are attributed to (111), (200), (220), (311) and (222) planes of NiO phase whereas the peaks diffracting at 44.42°, 51.76° and 76.26° represent (111), (200) and (220) planes of Ni phase. Both the phases crystallize in *Fm-3m* space group [38,39]. In addition to these phases, significantly intense diffraction peaks indexed as (002) and (100) from *h*-BN phase having *P6$_3$mmc* space group is also observed in *h*-BN/Ni-NiO nanocomposites, suggesting successful grafting of Ni-NiO nanostructures in the BN layers [40,41]. No impurity phase was found in these nanostructures/nanocomposites at least to the detection limit of the XRD. The intensity of the NiO peaks is relatively higher in case of nanocomposites when compared with the nanostructures. The increased percentage of the NiO phase in nanocomposites may be attributed to the enhanced strain induced by 2D *h*-BN sheets (*vide infra*). Another important piece of information which can be extracted from diffraction pattern is the increased peak width in case of NiO phase compared to Ni phase suggesting smaller crystallite size for NiO phase. Crystallite size (D) was calculated by Debye-Scherrer equation using the peak broadening of XRD peaks after correcting from the instrumental broadening (for details refer to Supporting Information). The crystallite sizes for the Ni phases were found to be around 27 nm whereas the size for the NiO phase is around 18 nm.

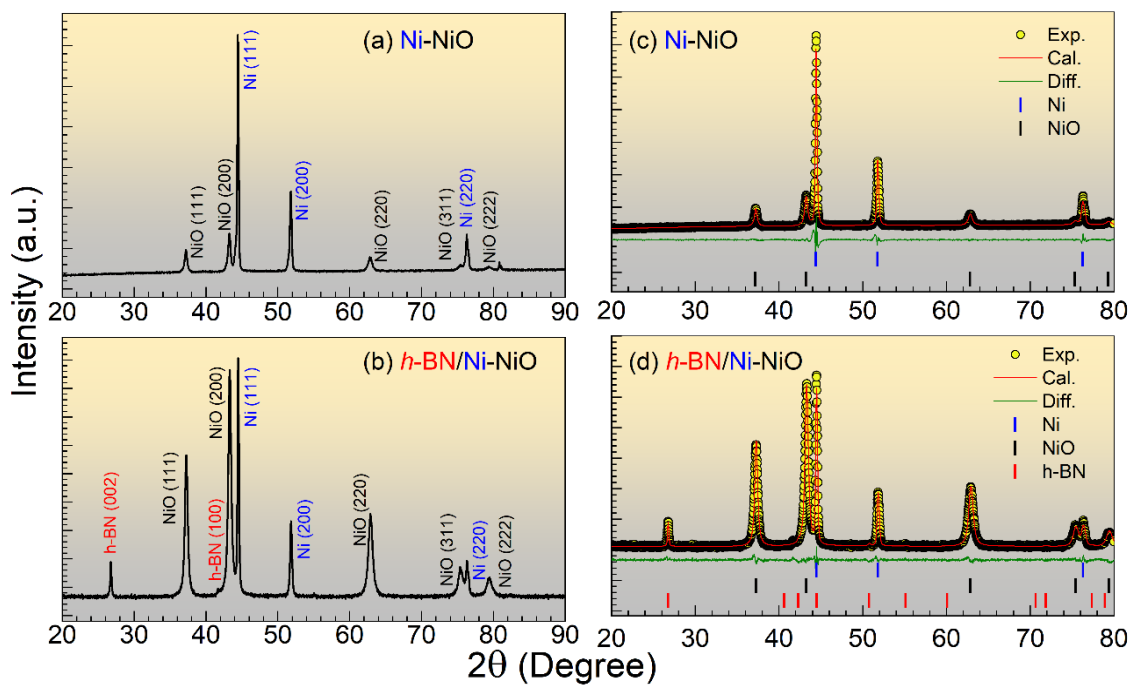

**Fig. 1.** Room temperature XRD spectra for (a) Ni-NiO nanostructures and (b) *h*-BN/Ni-NiO nanocomposites using Cu- Kα radiation where the miller indices for Ni, NiO and *h*-BN are written in blue, black and red color respectively. Rietveld refinements of XRD spectra for (c) Ni-NiO nanostructures (d) *h*-BN/Ni-NiO nanocomposites. Yellow colored circles, solid red lines, black and olive bars represent the experimental points, theoretical fit according to the structure and Bragg positions, respectively. The residuals are plotted at the bottom in olive color.

For better understanding of the crystal structure and quantitative measurements of phase formations, lattice parameters and microstructure, Rietveld refinements were performed using GSAS-II software package [42]. Different lattice parameters extracted from the refinements are given in Table S1. We observed that the best fit (as shown in Fig. 1(c)-(d)) can be obtained by considering *Fm-3m* space group of face centered cubic (fcc) structure of both Ni and NiO whereas *h*-BN is found to be in the *P6$_3$mmc* space group. Lattice parameters for all three phases are in accordance with known values. XRD data was further analyzed in form of Williamson-Hall plot [43] to extract the information about the crystallite size and strain (for details refer to Supporting Information). Upon integration with 2D *h*-BN, the strain is found to be higher than Ni-NiO nanostructures which could be attributed to the effect of intercalation and surface interaction of *h*-BN and Ni-NiO (Fig. S1).

The morphology of *h*-BN/Ni-NiO nanostructures was investigated using electron microscopic technique and the results are displayed in Fig. 2. SEM images reveal the distribution of nanosized particles decorated over a porous 2D surface (Fig. 2a). TEM images show that the aggregated spherical Ni-NiO particles are uniformly dispersed on the *h*-BN surface (Fig. 2b). The size distribution of Ni-NiO nanoparticles was analyzed using *ImageJ* software, and the average particle diameter was determined to be ~12 nm (Fig. S2). HRTEM analysis reveals lattice fringes corresponding to the (111) plane of NiO, with a *d*-spacing of 0.243 nm (Fig. 2c) [44]. The SAED pattern indicates the polycrystalline nature of Ni-NiO nanoparticles, with diffraction rings indexed to the (111), (200), (220) and (311) planes of NiO phase, consistent with PXRD and Rietveld refinement results (Fig. 2d). Additionally, EDS mapping confirms the presence of Ni, B, N, and O as the primary elements, with no detectable impurities (Fig. S3). The EDAX spectra further verify the elemental composition of the nanocomposite, confirming the presence and relative atomic and weight percentages of Ni, B, N, and O atoms (Fig. S4).

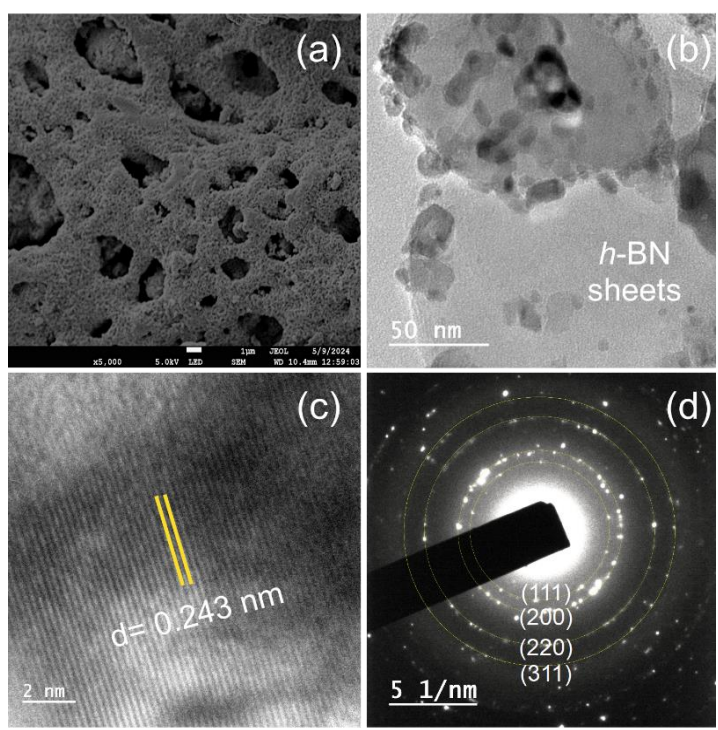

**Fig. 2.** Electron microscopic analysis *h*-BN/Ni-NiO. (a) SEM image; (b) TEM micrograph; (c) high resolution TEM displaying lattice fringes and (d) SAED pattern.

FTIR spectroscopy was employed to identify the functional groups and bonding characteristics of the *h*-BN/Ni-NiO nanocomposites (Fig. S5). A peak observed around ~ 473 cm$^{-1}$ corresponds to the Ni–O stretching vibration, indicating the formation of the NiO phase [45]. Additional bands centered at approximately 1024 cm$^{-1}$, and 1270 cm$^{-1}$ can be attributed to C–N stretching, and COO$^-$ bending vibration modes, respectively [45]. A distinct band at ~1630 cm$^{-1}$ is associated with the carboxyl stretching vibration, while a broad absorption band around 3440 cm$^{-1}$ and 3268 cm$^{-1}$ corresponds to O–H and N-H stretching vibrations. These vibrational features confirm the functionalization of glycine in Ni-NiO nanoparticles. Furthermore, the incorporation of *h*-BN into the composite is confirmed by the presence of its characteristic IR peaks. Strong peaks at 810 cm$^{-1}$ and 1381 cm$^{-1}$ are assigned to the out-of-plane B–N–B bending vibration and the in-plane B–N stretching vibration, respectively [46]. These FTIR results clearly demonstrate the successful formation of the $h$-BN/Ni-NiO nanocomposite.

UV-Vis spectra of Ni-NiO and $h$-BN/Ni-NiO nanocomposites were recorded in ethanol spanning from 200 to 600 nm range, as shown in Fig. S6. Ni-NiO exhibits an absorption band centered at 280 nm, corresponding to a direct band gap of 3.8 eV. In contrast, $h$-BN/Ni-NiO nanocomposite reveals two distinct absorption edges corresponding to optical transitions at 2.89 eV and 5.1 eV. These values are in good agreement with the reported band gaps of NiO and $h$-BN, respectively [47,48]. This decrease in band gap for the $h$-BN/Ni-NiO nanocomposite may be attributed to the electrostatic interactions between $h$-BN and Ni-NiO. Furthermore, we have performed photoluminescence (PL) measurements were conducted and compared the photoluminescence spectra of the $h$-BN/Ni-NiO composite system and the pure Ni-NiO (Fig. S7). A remarkable quenching of luminescence intensity was observed in the composite. This behavior strongly supports the occurrence of enhanced electron transfer in $h$-BN/Ni-NiO due to an improved charge separation facilitated by the heterointerface.

XPS analysis was performed to investigate the valence states and surface composition of the elements present in the as-prepared $h$-BN/Ni-NiO composites. The survey spectra (Fig. S8) confirmed the presence of Ni, C, B, N, and O as the main elements. High-resolution Ni 2p spectra exhibited two spin–orbit peaks at binding energies of 872.2 eV (Ni $2p_{1/2}$) and 853.9 eV (Ni $2p_{3/2}$). These peaks were deconvoluted using a Lorentzian function, revealing sub-peaks corresponding to Ni(0) and Ni(II) oxidation states, along with their respective satellite features (Fig. 3a) [49]. As shown in Fig. 3b, the high-resolution O 1s spectrum displayed an asymmetric profile, which could be deconvoluted into three distinct peaks. These were assigned to lattice oxygen (529.4 eV), adsorbed water molecules (531.1 eV), and surface-adsorbed or defect-related oxygen species (532.8 eV). The B 1s spectrum showed a peak at 190.2 eV, corresponding to B-N bonding (Fig. 3c) [50]. Furthermore, the deconvolution of the N 1s spectrum yielded a single peak at 397.8 eV, indicating a uniform chemical environment around nitrogen atom (Fig. 3d). A noticeable shift in binding energy for Ni $2p_{3/2}$, and lattice oxygen (O 1s) could be observed compared to the individual Ni-NiO nanocomposite as reported earlier [51]. These shifts indicate the presence of strong interfacial interactions between $h$-BN and Ni-NiO, which may contribute to the improved electronic and catalytic properties of the nanocomposite.

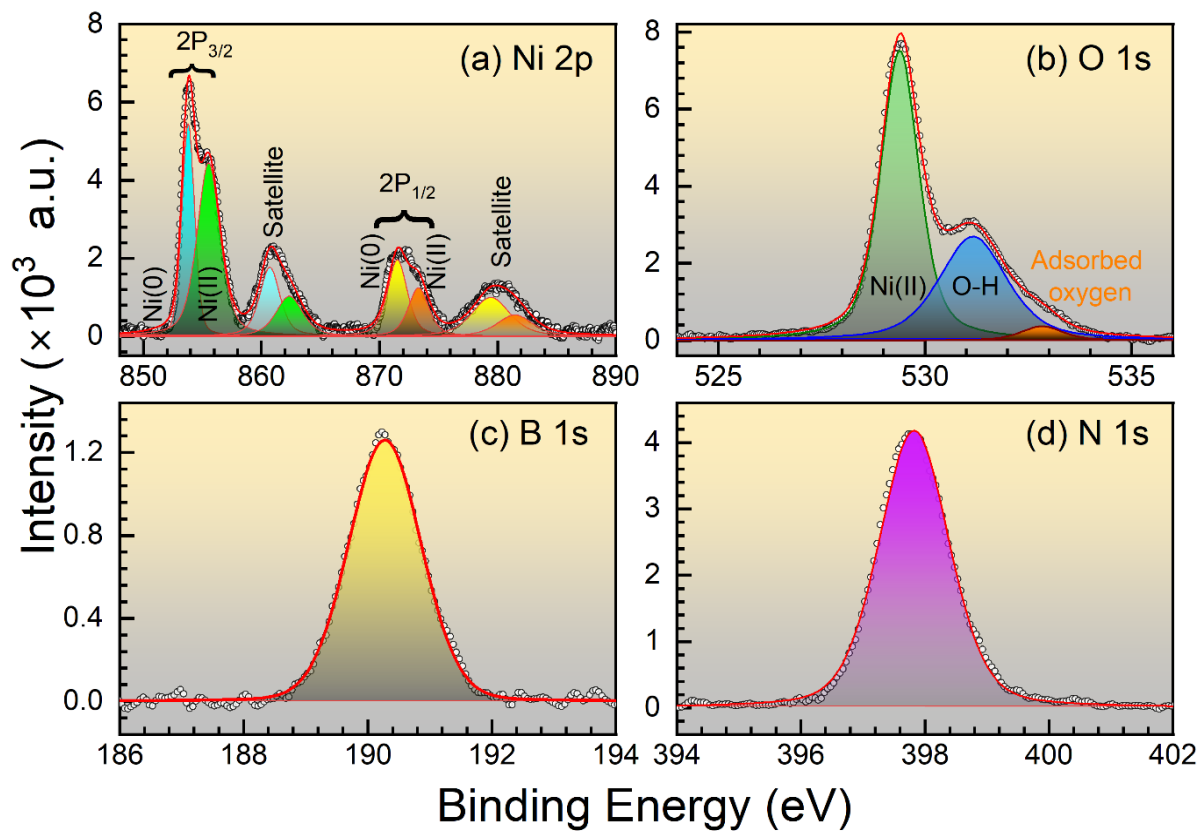

**Fig. 3.** Deconvoluted XPS spectra of (a) Ni 2p, (b) O 1s, (c) B 1s and (d) N 1s.

The surface area of the catalyst plays a crucial role in catalytic reactions, as it directly influences adsorption of the reactant molecules. Generally, a higher surface area offers more active sites for adsorption, thereby greater interactions among the reactant molecules, leading to enhanced catalytic efficiency and increased reaction rate [52]. To evaluate the surface area and porous characteristics of the synthesized nanocatalysts, Brunauer–Emmett–Teller (BET) analysis was conducted. Fig. 4 displays the nitrogen adsorption-desorption isotherm of *h*-BN/Ni-NiO. The isotherm exhibits a hysteresis loop characteristic of a type IV isotherm, which is indicative of mesoporous structures. The specific surface area of *h*-BN/Ni-NiO was measured to be 1269 $m^2/g$, which is significantly greater than that of Ni-NiO (4.033 $m^2/g$). Additionally, pore size distribution analysis using the Barrett–Joyner–Halenda (BJH) method revealed a pore volume of 20.4 $cm^3/g$ for *h*-BN/Ni-NiO, which is significantly higher compared to 0.0392 $cm^3/g$ for Ni-NiO. These findings suggest that the incorporation of *h*-BN markedly enhances both the surface area and pore volume of the Ni-NiO catalyst. Consequently, the *h*-BN/Ni-NiO composite is expected to demonstrate superior catalytic performance due to its improved surface properties as indicated by the BET measurements.

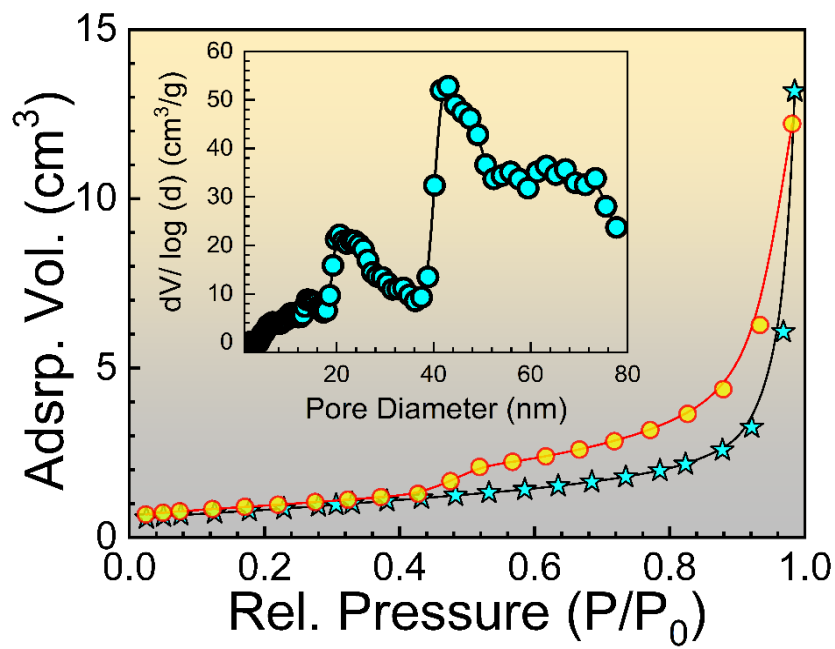

**Fig. 4.** BET nitrogen adsorption-desorption isotherms and pore size distribution plot (inset) of the *h*-BN/Ni-NiO nanocomposite.

**Catalytic reduction of nitrophenols**

To assess the catalytic performance of the nanocomposites in the reduction of nitro compounds, the model reaction involving the conversion of *p*-nitrophenol (*p*-NP) to *p*-aminophenol (*p*-AP) was selected. Although this transformation is thermodynamically favorable in the presence of NaBH$_4$ due to the difference in reduction potentials ($E_{p-NP/p-AP} = -0.76$; $E_{BH_4^-/H_3BO_3} = -1.33$ V), it is kinetically restricted [53]. Upon the addition of NaBH$_4$ to a solution of *p*-NP, the characteristic n→π* absorption band at 317 nm shifts to 400 nm, accompanied by a visible color change from yellow to green-yellow (Fig. 5). This shift is attributed to the formation of the nitrophenolate ion in the alkaline medium provided by NaBH$_4$. Notably, the absorbance of this 400 nm band remains unchanged over time in the absence of a catalyst, indicating no progress in the reduction reaction. However, upon introducing the *h*-BN/Ni-NiO catalyst, a rapid decrease in the 400 nm absorption band was observed, along with the emergence of a new band at 300 nm, characteristic of the n→π* transition in *p*-AP (Fig. 5). Since the decrease in absorbance at 400 nm was more prominent than the increase at 300 nm, the reaction kinetics were evaluated by tracking the decline of the 400 nm band (Fig. 6). Different amounts of catalyst were tested in reduction of nitrophenols using water as reaction medium to optimize the catalytic loading (Fig. S9). The concentration of NaBH$_4$ was optimized by varying its amount between 1 and 7 mg. A systematic increase in the rate of 4-nitrophenol reduction with increasing concentrations of NaBH$_4$ was observed (Fig. S10). This behavior supports the role of surface-adsorbed nickel-hydride intermediate in promoting hydride transfer to nitro groups.

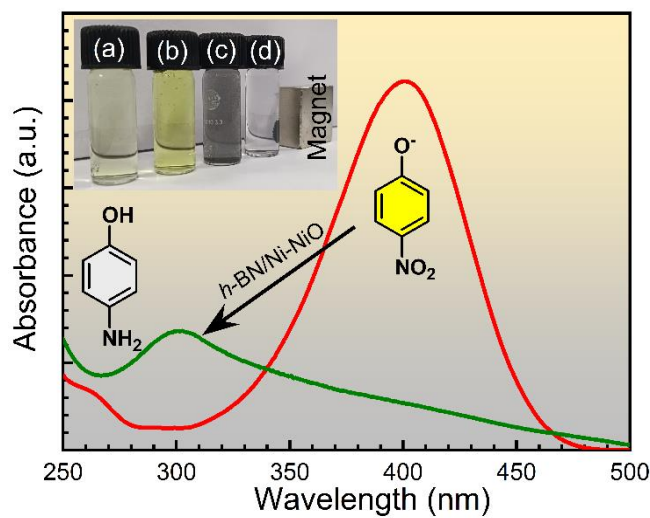

**Fig. 5.** UV-Vis spectral changes for the reduction of *p*-nitrophenolate to *p*-aminophenol. Inset: aqueous solution of (a) *p*-NP (b) *p*NP + NaBH$_4$ (c) *p*-NP + NaBH$_4$ + *h*-BN/Ni-NiO (d) after magnetic separation of *h*-BN/Ni-NiO.

To examine the role of *h*-BN incorporation in the Ni-NiO nanostructure, the catalytic reduction of two nitrophenol derivatives (*p*-NP and *o*-NP) was carried out using an excess of NaBH$_4$ (~100 equiv.) in the presence of three catalysts: *h*-BN/Ni-NiO, Ni-NiO, and *h*-BN alone. Under identical experimental conditions with 10 mg of catalyst dosage, both *h*-BN/Ni-NiO and Ni-NiO facilitated a rapid decrease in the absorbance of *p*-NP at 400 nm over time, along with a simultaneous increase in absorbance at 300 nm. This spectral shift indicates the successful reduction of *p*-NP to *p*-AP. The appearance of isosbestic points further supports a clean conversion pathway with no byproducts. Visually, the reaction mixture changed from a green-yellow to nearly colorless, supporting this transformation. Notably, the *h*-BN/Ni-NiO nanocomposite demonstrated significantly enhanced catalytic performance, completing the reduction in 3 min, whereas Ni-NiO required 12 min (Fig. 6 a-b). In contrast, *h*-BN alone showed negligible activity even after 20 min (Fig. S11). A similar trend was observed for the reduction of *o*-NP. Using *h*-BN/Ni-NiO as the catalyst, the reaction reached near-complete reduction within 6 min, whereas Ni-NiO required 25 min to achieve ~ 95% conversion (Fig. 6 c-d). These results emphasize the synergistic effect of *h*-BN and Ni-NiO heterojunction in enhancing the catalytic reduction efficiency of the *h*-BN/Ni-NiO system.

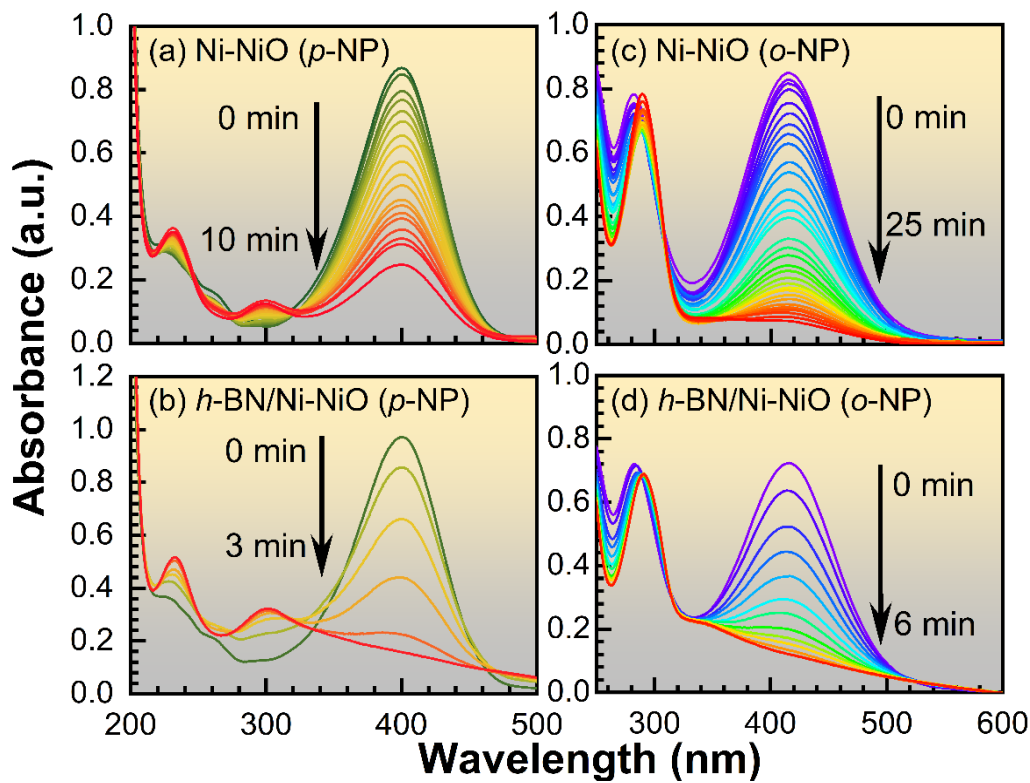

**Fig. 6.** Time-dependent UV-Vis spectral changes for catalytic reduction of nitrophenols using (a, c) Ni-NiO and (b, d) *h*-BN/Ni-NiO as catalyst.

In the reduction process of nitrophenols, the concentration of NaBH$_4$ (~2.5 mM) is approximately 50 times greater than that of *p*-NP (0.05 mM). As a result, NaBH$_4$ is present in excess throughout the reaction, allowing the reduction to be treated as a pseudo-first-order kinetic reaction with respect to nitrophenol. The reaction kinetics can be described by the equation: $\ln(C_0/C) = kt$, where $C_0$ and $C$ represent the concentrations of nitrophenolate ions at time zero and at time t, respectively, and k is the apparent rate constant. The plot of $\ln(C_0/C)$ versus time (t) showed a good linear correlation, confirming the pseudo-first-order behavior. The calculated rate constants for *p*-NP were found to be 0.573 min$^{-1}$ for *h*-BN/Ni-NiO, 0.115 min$^{-1}$ for Ni-NiO, and 0.018 min$^{-1}$ for *h*-BN (Fig. 7a). Similarly, for the reduction of *o*-NP, the rate constants were 0.373 min$^{-1}$ for *h*-BN/Ni-NiO, 0.118 min$^{-1}$ for Ni-NiO, and 0.017 min$^{-1}$ for *h*-BN (Fig. 7b). These results clearly demonstrate that the catalytic efficiency follows the order: *h*-BN/Ni-NiO > Ni-NiO >> *h*-BN. It is evident that the *h*-BN/Ni-NiO catalyst demonstrated excellent catalytic performance compared to previously reported NiO-based catalysts and even many noble metal-supported systems [54–57]. As control experiment *p*-nitrophenol reduction reaction was performed using hydrazine as an alternative electron donor and h-BN/Ni-NiO as catalyst under identical conditions. A negligible reduction of *p*-nitrophenol in the presence of excess hydrazine was observed (Fig. S12). This underscores the unique efficiency of BH$_4^-$ in facilitating electron transfer and highlights its critical role in the catalytic mechanism.

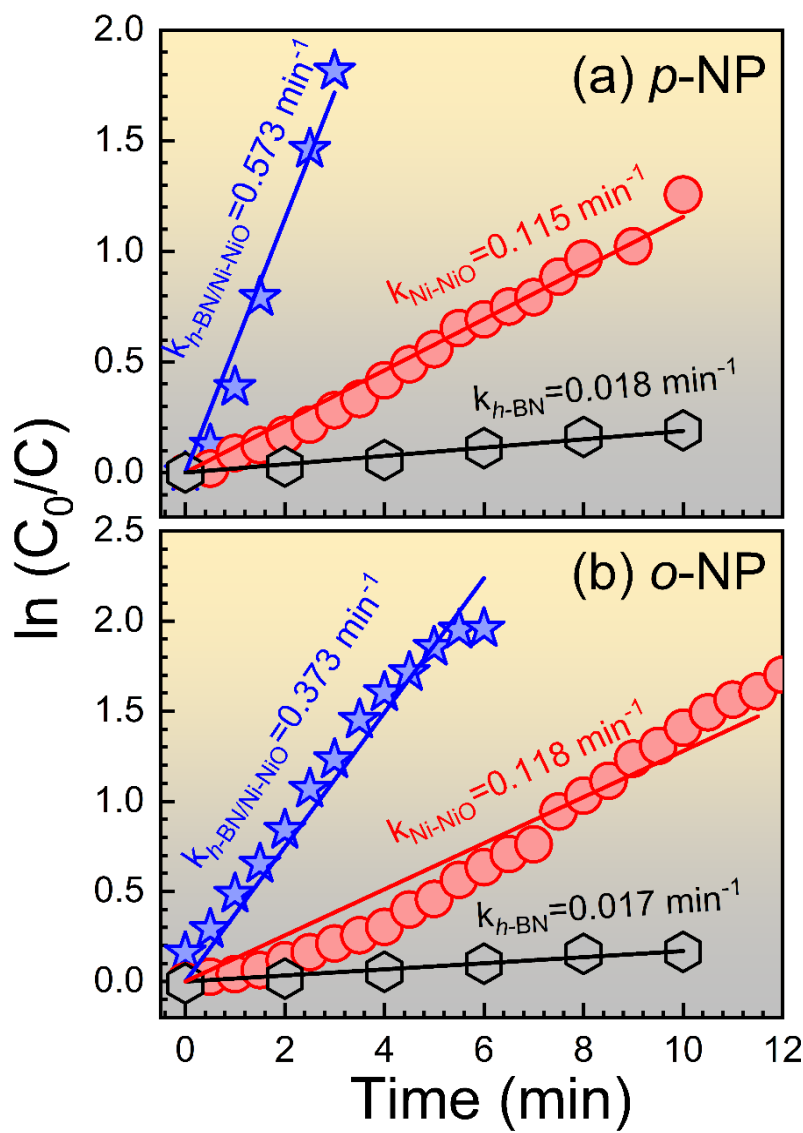

**Fig. 7.** Plots of ln(C₀/C) versus time for the catalytic reduction of (a) *p*-NP and (b) *o*-NP using the as-prepared catalysts.

Reusability and easy separation of catalysts are crucial factors for large-scale and sustainable chemical conversions. Stability and reusability studies were performed by magnetically separating the catalyst from reaction mixture and reusing it for 6 cycles. Even after the 6$^{th}$ cycle, *h*-BN/Ni-NiO catalyst displayed minimal change in catalytic activity which indicates the high stability and reliability of this catalytic system (Fig. S13). The improved stability could be attributed to the well dispersed Ni-NiO nanostructures on to the *h*-BN nanosheets. Intercalation and strong interaction between Ni-NiO and *h*-BN may provide a good stability to the composite structure. The porous structure with large surface area provides abundant active sites of the catalysts and it also facilitates the desorption of *p*-AP from the catalyst surface.

**Plausible mechanism**

The catalytic mechanism for the reduction of nitrophenol derivatives generally involves four key steps: (i) adsorption of reactants onto the catalyst surface, (ii) diffusion of active species, (iii) hydrogenation of the nitro group (-$NO_2$), and (iv) desorption of the resulting product. The overall reaction rate is primarily governed by the electron transfer from the electron donor ($BH_4^-$) to the electron acceptor (NP), which depends heavily on the proximity and interaction between the two species. Initially, $BH_4^-$ ions adsorb onto the catalyst surface and react with water to generate $BO_2^-$ and molecular hydrogen ($H_2$) in the aqueous phase [58]. A significant portion of this $H_2$ is adsorbed onto the surface of *h*-BN/Ni-NiO, forming Ni-hydride complexes that act as the active reductants. Simultaneously, nitrophenolate anions are adsorbed onto the catalyst surface *via* both physisorption and chemisorption, particularly through interactions between the nitro group and surface metal or nitrogen atoms of *h*-BN (Scheme-2). This close spatial arrangement brings the substrate in proximity to the active reductant. Subsequently, electrons are rapidly transferred from the hydride species through the *h-BN* surface to the adsorbed nitrophenolate ions, thereby accelerating the reduction of NP to AP. Finally, the AP product desorbs from the catalyst surface, allowing the cycle to continue. The enhanced catalytic activity and stability of the *h*-BN/Ni-NiO catalyst can be attributed to several key factors: (i) the mesoporous *h*-BN support with uniformly dispersed Ni-NiO nanoparticles offers a high surface area and abundant active sites, facilitating efficient electron transport from oxidation to reduction sites; (ii) metal-metal oxide and *h*-BN/Ni heterojunctions interfaces enhance charge transport and reaction kinetics; (iii) *h*-BN contributes to the improved electronic conductivity and provides coordination sites that stabilize Ni-NiO nanoparticles, boosting both reactivity and stability and (iv) the polarized nitrogen atoms in *h*-BN can attract hydroxyl or nitro groups from nitrophenolate ions, enhancing the adsorption and orientation of reactants on the catalyst surface.

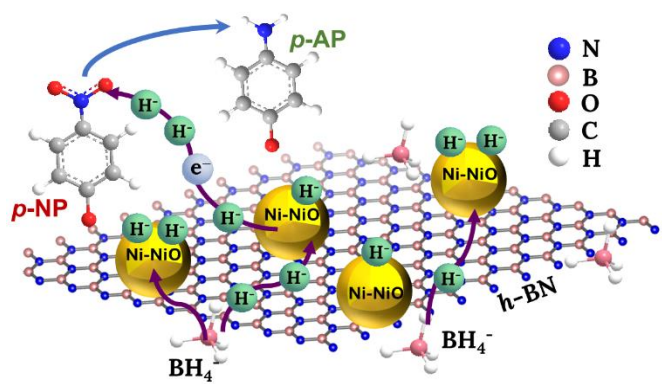

**Scheme 2.** Possible mechanism for the catalytic reduction of *p*-NP.

## 4. Conclusions

In summary, this study successfully demonstrated the synthesis of magnetic Ni-NiO nanostructures and their *h*-BN/Ni-NiO composites *via* a simple and scalable combustion method. Spectro-analytical characterization elucidated the phase composition and uniform dispersion of the nanoparticles within the *h*-BN matrix. The *h*-BN/Ni-NiO nanocomposite showed superior catalytic efficiency in the reduction of nitrophenol derivatives, highlighting the synergistic interaction between *h*-BN and Ni-NiO. Additionally, the magnetic properties of the composite facilitated easy recovery and reuse, maintaining high catalytic activity over 6 cycles. These findings position the *h*-BN/Ni-NiO nanocomposite as a highly effective, reusable, and environmentally friendly catalyst, with significant potential for practical applications in green chemistry and environmental remediation.

## Acknowledgements

A. V. is thankful to UPES, Dehradun, India for providing a PhD fellowship. S. L. sincerely acknowledge SEED (Grant No.: UPES/R&D-SOE/20062022/02) and SEED-INFRA (Grants No.: UPES/R&D-SEED-INFRA/24082023/04 and UPES/R&D-SoAE/08042024/20) research funding from UPES, Dehradun, India. Authors are thankful to the CIC, UPES Dehradun, for extending research infrastructure.